\documentclass[nofootinbib,amsfonts,prd,aps]{revtex4}
\usepackage{subfig}
\usepackage{amsmath,amssymb,amsfonts,amsthm,mathrsfs}
\usepackage{stackrel}
\usepackage{graphicx}
\usepackage{nameref}

\begin{document}

\title{Critical collapse of a scalar field in semiclassical loop
  quantum gravity}

\author{Florencia Ben\'{\i}tez$^{1}$, Rodolfo Gambini$^{1}$, Luis
  Lehner$^{2}$, Steve Liebling$^{3}$, Jorge Pullin$^{4}$}
\affiliation {
1. Instituto de F\'{\i}sica, Facultad de Ciencias, 
Igu\'a 4225, esq. Mataojo, 11400 Montevideo, Uruguay. \\
2. Perimeter Institute, 31 Caroline St, Waterloo, ON N2L 2Y5,
Canada.\\
3. Long Island University, Brookville, NY 11548.\\
4. Department of Physics and Astronomy, Louisiana State University,
Baton Rouge, LA 70803-4001}

\begin{abstract}
We study the collapse in spherical symmetry of a massless scalar
field minimally coupled to gravity using the
semiclassical equations that are expected from loop quantum
gravity. We find
critical behavior of the mass as a function of the parameters of the
initial data similar to that found by Choptuik in classical general
relativity for a large set of initial data and values of the
polymerization parameter.  Contrary to wide expectations for quantum gravity, our semiclassical
field equations have an exact scale invariance, as do the classical
field equations. As one would then expect, we numerically find that
the phase transition is second order, again as in the classical case.
\end{abstract}

\maketitle
Choptuik \cite{choptuik} studied numerically the collapse of a
massless, minimally coupled, scalar
field coupled to general relativity. For a one parameter family of
initial data he noted that there exists a critical value of the
parameter. Below it, the scalar field disperses to infinity. Above it,
a black hole forms through a second order phase transition. The dependence of the final mass of the black hole
on the parameter of the initial data has a universal form $M_{\rm BH}\sim
\left(p-p_*\right)^\beta$ where $p_*$ is the critical value and
$\beta\sim 0.37$  is a universal exponent, independent of the choice
of parameter and initial data, provided $p_*$ is non-vanishing. This
critical behavior and universal scaling has been observed for several
other systems (see \cite{gundlach} for a review).  While it seemed likely 
that the transition would be second order as there was no natural length scale in the
problem, before these numerical studies the order of the phase transition  
was unsettled \cite{christodoulou}.
This opens the question of how things could change in a
quantum treatment of the collapse. Quantum gravity has a natural
length scale, the Planck length. Indeed, previous studies of
polymerized dynamics of metric general relativity seemed to suggest
that the transition becomes first order \cite{husain}. Even today, a
complete quantum treatment of the problem is not available.

Here we study the critical collapse of massless scalar fields,
minimally coupled to the
semi-classical equations that stem from loop quantum gravity with
spherical symmetry \cite{gambiniolmedopullin}. In it, the classical
variables for gravity are given by a the spherical remnants of the triads in the radial and
transverse directions $E^x$ and $E^\varphi$ and their canonically
conjugate momenta
$K_x$ and $K_\varphi$. The
metric of space-time can be written as,
\begin{equation}
ds^{2}=-\alpha^{2}dt^{2}+\varLambda^{2}dr^{2}+R^2d\Omega^{2}
\end{equation}
and the relation to the loop quantum gravity triads are
$\Lambda=E^\varphi/r$, $R^2=\vert E^x\vert$ and to the extrinsic
curvatures $K_{xx}=-{\rm sign}(E^x) \left(E^\varphi\right)^2
K_x/\sqrt{\vert E^x\vert}$ and $K_{\theta \theta}= -\sqrt{\vert
  E^x\vert} A_\varphi/(2\gamma)$ with $\gamma$ the Immirzi parameter.

To try to stay as close to Choptuik's treatment as possible, we 
choose coordinates such that  $E^x=r^2$. This corresponds to the usual
Schwarzschild radial coordinate and eliminates $K_x$ through the diffeomorphism
constraint (one of the Einstein equations).  His ``polar condition'' ($K^r_r={\rm Tr}(K)$) corresponds in these
variables to $K_{\varphi}=0$, which makes the metric diagonal. This has the
unexpected effect of making the gravitational part of the
semiclassical equations reduce to the classical form (both
$K_{\varphi}$ and $K_x$ drop out from the equations and these would be
the variables that would get polymerized in the semi-classical
theory). The only effect of the loop quantization is in the
polymerization of the scalar variables.  The need to consider polymeric
representations for scalar fields in loop quantum gravity was first
pointed out by Thiemann \cite{thiemann} as a need to deal with
diffeomorphism invariance and have a well defined measure in the space
of matter fields. It might be possible that in a more complete
treatment using different coordinates, effects from polymerization of
the gravitational variables could potentially produce somewhat
different results than those of this paper.

The system of classical
equations  (for $G=c=1$) is \cite{saeed},
\begin{equation}
\frac{\alpha'}{\alpha}-\frac{\left(E^{\varphi}\right)'}{E^{\varphi}}+\frac{2}{r}-\frac{\left(E^{\varphi}\right)^{2}}{r^{3}}=0\label{eq:5}
\end{equation}

\begin{equation}
\frac{\left(E^{\varphi}\right)'}{E^{\varphi}}-\frac{3}{2r}+\frac{\left(E^{\varphi}\right)^{2}}{2r^{3}}-2\pi r\left(\Pi^{2}+\Phi^{2}\right)=0\label{eq:6}
\end{equation}

\begin{equation}
\dot{\Phi}=\left(\frac{\alpha r}{E^{\varphi}}\Pi\right)^{'}\label{eq:7}
\end{equation}

\begin{equation}
\dot{\Pi}\text{=}\frac{r\alpha}{E^{\varphi}}\Phi'+\left(\frac{r\alpha'}{E^{\varphi}}+\frac{3\alpha}{E^{\varphi}}-\frac{r\alpha\left(E^{\varphi}\right)'}{\left(E^{\varphi}\right)^{2}}\right)\Phi\label{eq:8}
\end{equation}
where $\Phi\equiv\phi'$ and
$\Pi\equiv\frac{E^{\varphi}}{\alpha r}\dot{\phi}$. The first equation
determines the lapse ($\alpha$) and arises from imposing
$K_\varphi=0$. The second equation is the Hamiltonian constraint
(another of the Einstein equations). The last two equations are the
evolution equations (the rest of the Einstein equations).

To construct the semi-classical equations we polymerize the scalar
field $\phi\to\frac{\sin\left(k\varphi\right)}{k}$, and its canonical
momentum, $P_{\phi}\to P_{\varphi}$. A more detailed discussion of the
polymerization of scalar fields can be seen in
\cite{lewandowski}. This is also a construction that has been
extensively used in the context of loop quantum cosmology (see
\cite{assingh} for a review).  $k$ is the polymerization parameter. In
the cases in which the variable being polymerized is a connection (as
in the gravitational variables), its interpretation is associated with
the loops appearing in the holonomies of the loop representation.  The
fundamental discreteness that appears at the quantum level establishes
a lower bound for these parameters, of the order of the Planck
scale (in our case the parameter has dimensions of length so the
natural
scale would be the Planck length). Notice that in this context the most natural thing is to
polymerize the configuration variables, which in the gravitational
case are connections. In the context of polymerized metric theories it
is not clear which is a more natural choice, whether to polymerize the
configuration variable or the momentum \cite{husain}. Polymerizing is
not guaranteed to produce the correct semiclassical theory. In all
examples studied up to now it has.  Ideally one would derive a
semiclassical theory from a full quantum theory of gravity, but
unfortunately no such description is known even for this model. Instead, we settle for the
polymerized theory as the best candidate available for a semiclassical
theory.

The
resulting semiclassical equations become,
\begin{equation}
\frac{\alpha'}{\alpha}-\frac{\left(E^{\varphi}\right)'}{E^{\varphi}}+\frac{2}{r}-\frac{\left(E^{\varphi}\right)^{2}}{r^{3}}=0,\label{eq:13}
\end{equation}

\begin{equation}
\frac{\left(E^{\varphi}\right)'}{E^{\varphi}}-\frac{3}{2r}+\frac{\left(E^{\varphi}\right)^{2}}{2r^{3}}-2\pi r\left(\frac{\left(P_{\varphi}\right)^{2}}{r^{4}}+\left(\varphi'\right)^{2}\cos^{2}\left(k\varphi\right)\right)=0,\label{eq:14}
\end{equation}

\begin{equation}
\dot{\varphi}=\frac{\alpha}{E^{\varphi}r}P_{\varphi},\label{eq:15}
\end{equation}

\begin{equation}
\dot{P}_{\varphi}=\frac{r^{2}}{E^{\varphi}}\left[\left(\frac{3\alpha E^{\varphi}-r\alpha\left(E^{\varphi}\right)'+\alpha'E^{\varphi}r}{E^{\varphi}}\right)\varphi'\cos^{2}\left(k\varphi\right)+r\alpha\varphi''\cos^{2}\left(k\varphi\right)-r\alpha k\left(\varphi'\right)^{2}\cos\left(k\varphi\right)\sin\left(k\varphi\right)\right],\label{eq:16}
\end{equation}
and one recovers the classical limit when $k \to 0$, one has that
$\varphi$ reduces to $\phi$ and $P_\varphi$ to $r^2\Pi$ in that
limit. To facilitate comparison with Choptuik's notation it should be
noted that $E^\varphi=r a$ in terms of his variables. It should
be noted that the polymerized equations retain the scaling symmetry of
the classical equations $r\to c r$, $t\to c t$ with $c$ a constant. This is a key
difference with other polymerized treatments based on metric
variables \cite{husain} which introduce a length scale dependent correction near the origin
for the radial coordinate, and that found a mass gap. It should be noted that Garfinkle
\cite{garfinkle2} pointed out  that
this symmetry is a necessary condition for the existence of a self
similar critical
solution and therefore a zero mass gap. And as we shall see, indeed no
mass gap seems to develop.

\begin{figure}[h]
\includegraphics[height=5cm]{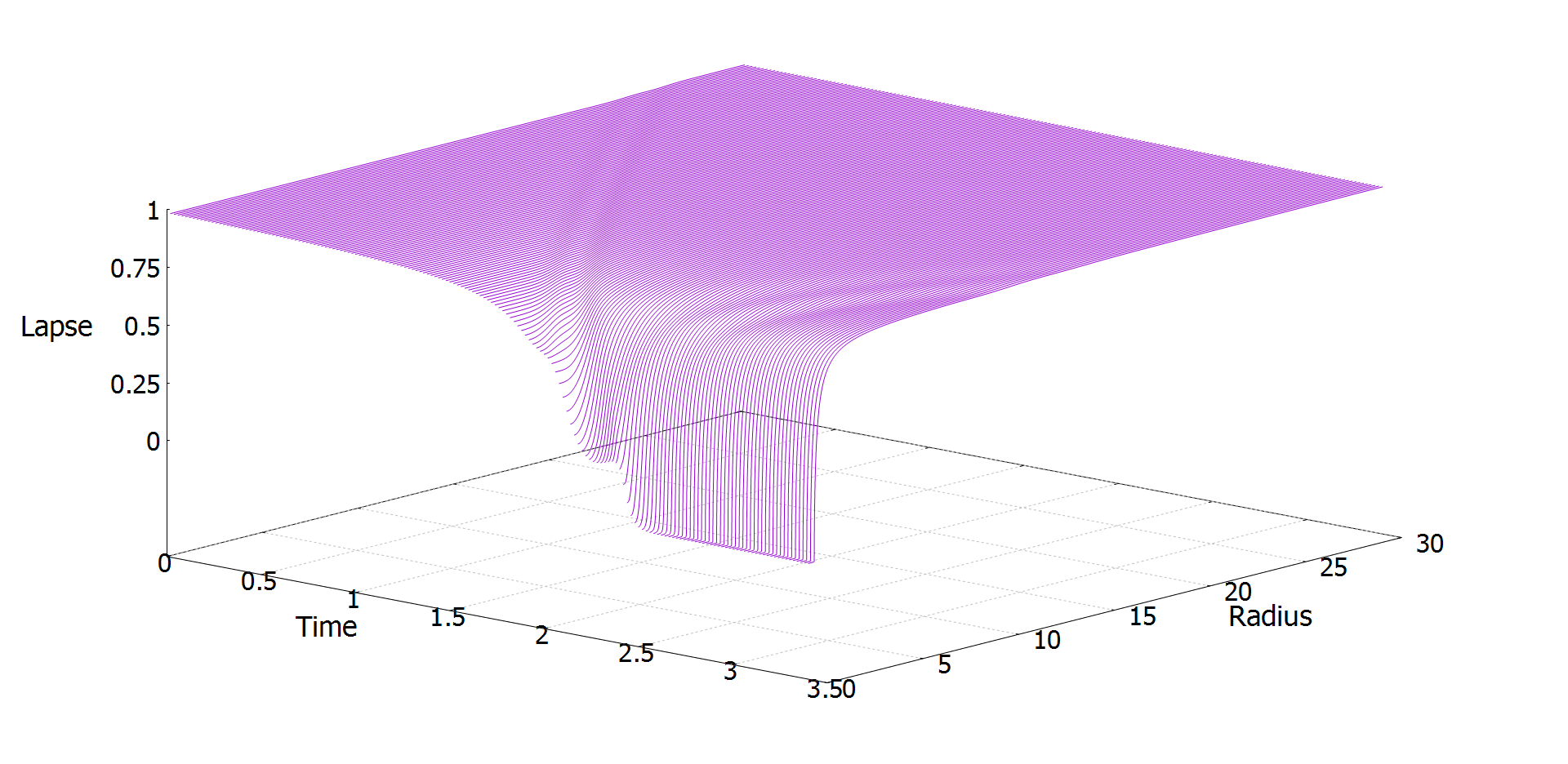}
\caption{
Demonstration of the collapse of lapse signifying black hole
formation.  The lapse deviates from one in the region of the initial
pulse which travels toward $r=0$. As the pulse encounters the origin of
coordinates, the lapse quickly decreases to zero indicating the
formation of a black hole. An outgoing wave of scalar field is seen
which is not captured by the black hole formation. For this evolution,
the polymerization parameter $k$ is unity, a value large given
expectations of the parameter being of the Planck scale, and yet the
dynamics is very similar to the classical case for which $k=0$.} 
\label{fig1}
\end{figure}

\begin{figure}[h]
\includegraphics[height=5cm]{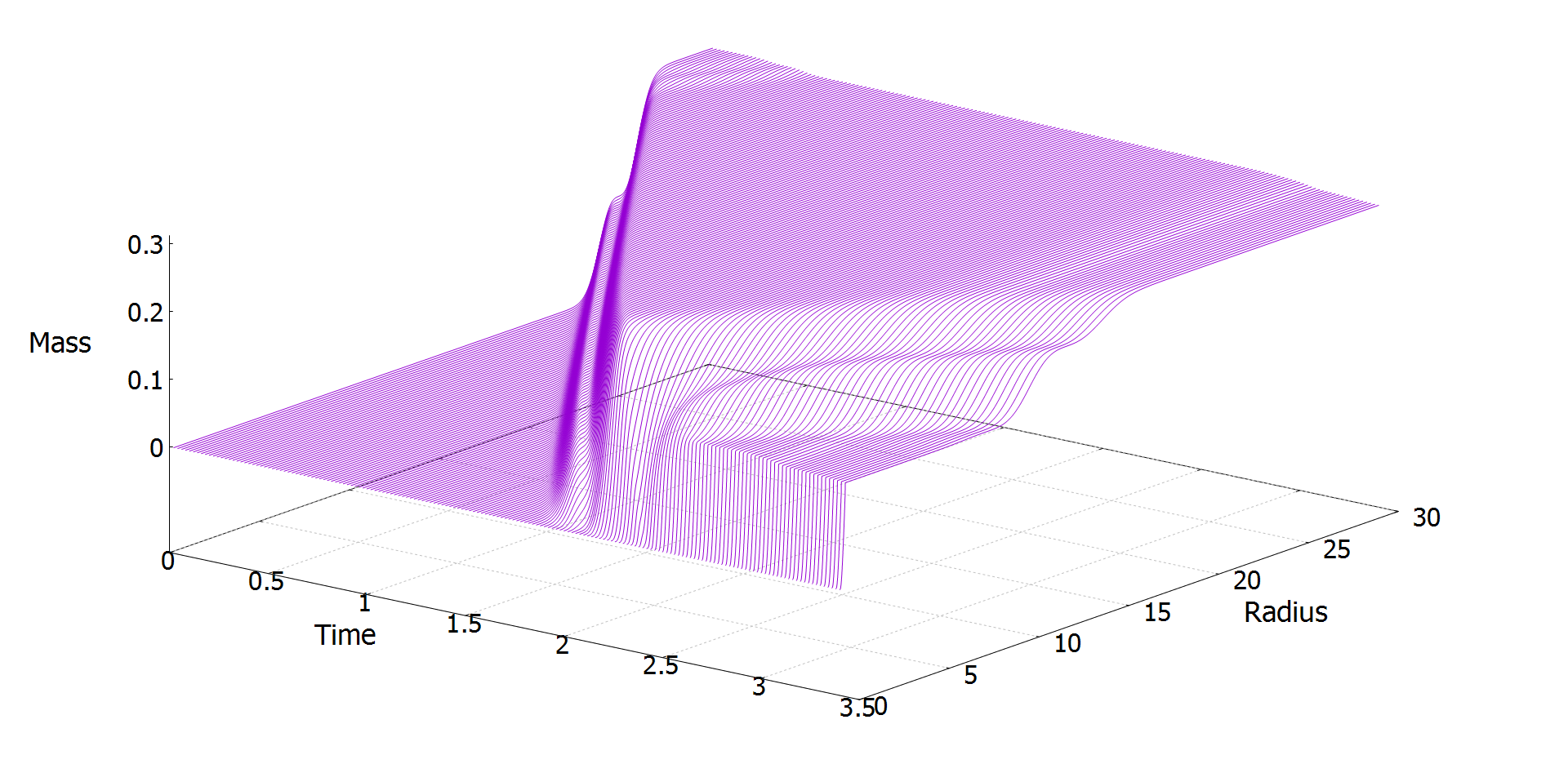}
\caption{The behavior of the mass aspect function $m(r,t)$ defined as
  $g_{rr}(r,t)=(1-2 m(r,t)/r)^{-1}$ for the same evolution as shown in
  figure 1. It quickly settles to its final
  value after the formation of the black hole.}
\label{fig2}
\end{figure}

We proceed to integrate the equations adapting a version of Choptuik's
original code (available publicly at \cite{choptuiklaplace}). This
paper can be seen as a first approach to the problem, in particular with
a few exceptions (to confirm the observed behavior) we have not used 
adaptive mesh refinement as in Choptuik's studies. We
choose as initial data family a set of Gaussians parameterized as
$\phi(r)= \phi_0\exp\left(-\left[(r-r_0)/\delta\right]^q\right)$. We
will keep $r_0=25, \delta=1.5$ and $q=2$ fixed and vary $\phi_0$,
which we call the parameter $p$. The simulations show that there is a
critical value $p_*$ of the parameter below which no black hole forms
and above it one sees the collapse of the lapse typical of the
formation of a black hole, as shown in figure \ref{fig1}. The
coordinate system we are using cannot penetrate the horizon. However,
there are clear indications of the formation of a black hole {\em in appropriate regimes}
(see further discussions below). For
instance the mass aspect $m(r,t)=(1-\Lambda^{-2})r/2$ is always smaller than $r/2$ and
tends to that value when the lapse vanishes, as shown in figure \ref{fig2}.  From this we can get an
approximation to the mass of the black hole.

To find the critical value $p_*$ we used a method of binary search in
which one increases monotonically the value of the parameter until a
black hole is formed, then one backtracks and brackets the critical
value. 

\begin{figure}[h]
\includegraphics[height=7cm]{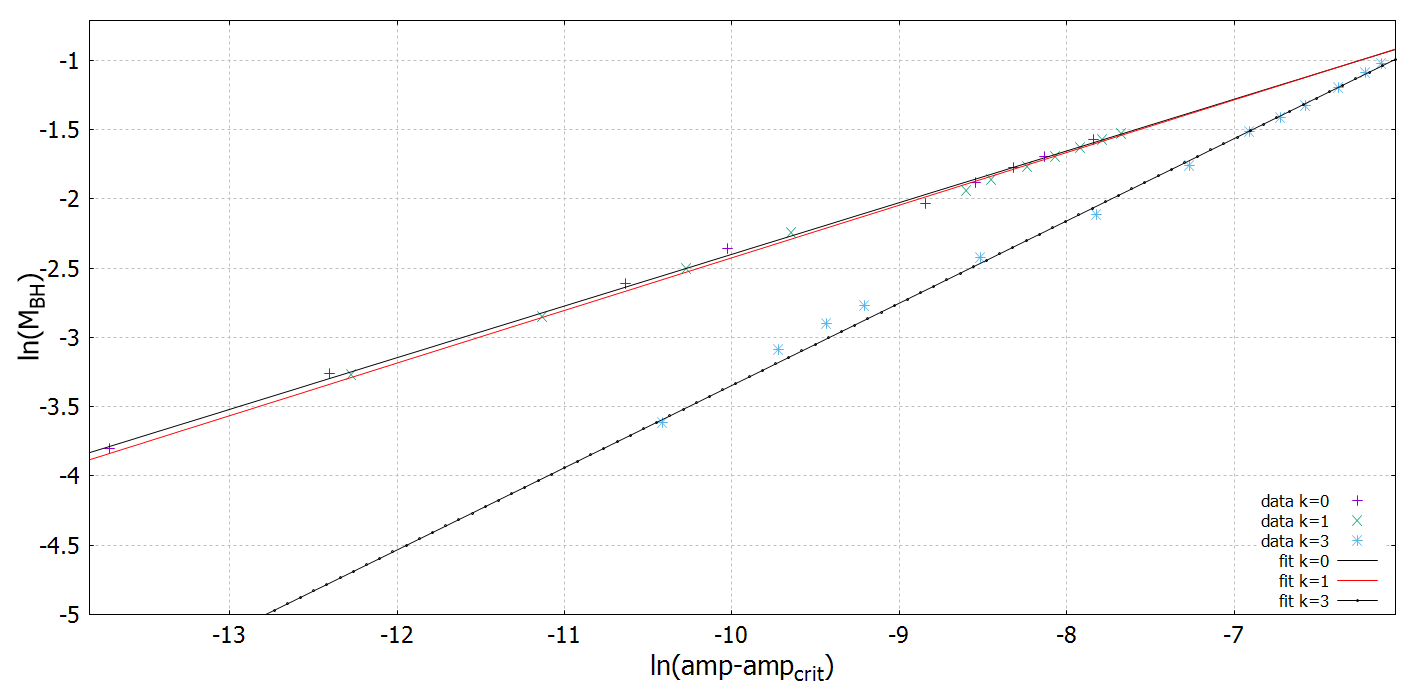}
\caption{Plots of the final black hole mass as a function of
  the parameter of the initial data for different values of the
  polymerization
  parameter $k$. They show the scaling observed by
  Choptuik, $M_{BH}= C (p-p_*)^\gamma$ where $C$
  depends on the value of the polymerization parameter
  $k$ and only a mild deviation 
from the value of $\gamma\sim 0.37$ as a function of
$k$. It should be recalled that $k$ is supposed to be Planck
scale, so values like $k=1$ are already quite exaggerated.
Plus signs correspond to $k=0$ (the case studied by Choptuik),
x corresponds to $k=1$, stars to $k=3$.}
\label{kp5}
\end{figure}

Figure \ref{kp5} shows the behavior of the final black hole mass as a
function of the parameter $p$, for various values of the
polymerization parameter. We see the same behavior Choptuik
encountered in the classical theory with the same universal exponent
 and a very mild dependence on the polymerization parameter (up to
$k=0.5$ the exponent remains the same, within numerical errors
---we  stress that physically this is an unrealistically large value---).
The mass scales as $M_{\rm BH}= C (p-p_*)^\gamma$ where $C$ depends on the
value of $k$ (and, as in the usual case, on the particulars of the
initial data) but we do not detect significant deviations from the
value of $\gamma\sim 0.37$ observed in the classical case unless we
force very large values of the polymerization parameter (it should be
remembered that it is supposed to be Planck scale).  We have
run tests with other families of initial data confirming its
universality.

An interesting point to be discussed is that the polymerized theory
has a maximum departure from the classical theory when
$k\varphi\sim \pi/2$. By observing simulations close to criticality but
sub-critical, we note that $k\varphi$ is always considerably smaller
than $\pi/2$ for a given $p$ in the domain covered (which in the case
of black hole formation is only the black hole exterior), even for
$k=1$, which as we have argued, is already an unrealistically large
value. What can be happening is that we are not getting close enough
to criticality and if one did, regions with $k\varphi\sim \pi/2$ might
occur at the origin, where one expects large curvatures to develop.
This is in line with the expectation that the polymerized theory will depart from general
relativity only close to where the singularity
was supposed to be. Solutions of the quantum theory for eternal black
holes reinforce this belief \cite{gapu}. More careful analysis,
perhaps with horizon penetrating coordinates and adaptive mesh
refinement, will be needed to confirm these points and others, like
the self-similar scaling seen in the classical case. We
will study this in future work using adaptive mesh refinement.

If one considers this model simply as a dynamical system, then it is
interesting to study the regime in which the product $k\varphi$
 becomes dynamically large (that is, close to $\pi/2$). It should be noted that 
the equations for the polymerized
scalar field allow for the formation of shocks/rarefaction as propagation
speeds depend on $\cos(k \varphi)$. Such phenomena requires additional conditions to
pick a unique solution and there is not a well developed  theory for handling them in this case. 
Certain initial choices of  $k\varphi$ can lead
to rather complex behavior in timescales shorter than potential black hole
formation developing features which are hard to follow even with adaptive mesh
refinement. Within families with large initial $k\varphi$ one may find
``islands'' where the behavior is similar to the one we observed for
small $k$'s and is analogous to the classical Choptuik behavior, but
that are surrounded by initial data that may not even form black
holes. More study is needed to understand the full phase space of
initial data when $k\varphi$ may be allowed to be large.
It also urges some caution to conclude things about the interior,
which our code cannot cover. It should also be noted that, although
potentially interesting from a mathematical point of view, solutions
with large $k\varphi$ are really beyond the realm of physical
applicability of the semiclassical theory we are considering. It is
well known that the Choptuik phenomenon close to criticality generates
large curvatures near the horizon, and it is widely expected that
large curvature regions require full quantum gravity for their
description. Although it has been observed, in the context of loop
quantum cosmology, that the semiclassical theory works well even in
the deep quantum regime~\cite{rovelli}, we have no reason to expect something
similar in our case. Nevertheless, the fact that propagation speeds are dependent
on the value of $\varphi$ when $k \neq 0$ could have potentially observable consequences
even in a regime where a semiclassical approach would apply. The potential reach of
this observation should be explored. Another point to be considered
is that we have considered a polymerization with a constant
parameter. In loop quantum cosmology at least, it has proven more
physically correct to use polymerization parameters that depend on the
dynamical variables \cite{assingh}. This issue has not been significantly explored
out of the cosmology context and may wish to be considered in future
analysis of the situation studied in this paper.

Summarizing, we have studied the critical collapse of a massless,
minimally coupled, scalar field
in a version of semiclassical, spherically symmetric loop quantum gravity. We find
that the results for the scaling of the mass agree with those of
classical general relativity with very mild dependence on  the
polymerization parameter and no mass gap (minimum value of the black hole mass). We plan
on carrying further studies of the echos that are present near the
critical solution in a forthcoming paper using adaptive mesh
refinement to see if the self-similarity observed in the classical
case persists. We would also like to probe whether the wiggles \cite{piran} that appear in the
exponent also appear. We also wish to probe closer to where the singularity
would be in the classical theory to see if the behavior observed there
of the curvature \cite{garfinkle} is present or is modified by the
polymerization. We would like to probe better the case in which
departures from the classical theory are large already at the level of
the initial data.

\section*{Acknowledgment}

We wish to thank Javier Olmedo for discussions. This work was
supported in part by Grants NSF-PHY-1603630, 1903799, 1827573 and
1912769, funds of the Hearne Institute for Theoretical Physics,
CCT-LSU, fqxi.org and Pedeciba.  LL was supported in part by NSERC,
and CIFAR.  FB thanks the Perimeter Institute for Theoretical Physics
for hospitality.  Research at Perimeter Institute is supported by the
Government of Canada and by the Province of Ontario through the
Ministry of Research, Innovation and Science.

\end{document}